\documentclass[runningheads]{llncs}
\usepackage[T1]{fontenc}
\usepackage{amsfonts}
\usepackage{mathtools}
\usepackage{amssymb}
\usepackage{array}
\usepackage{graphicx}
\usepackage{siunitx}
\usepackage{booktabs}
\usepackage{color}
\usepackage{hyperref}

\hypersetup{
  colorlinks,
  linkcolor={black},
  citecolor={black},
  urlcolor={blue},
}
\begin{document}
\title{Extending Liquid Rank Toward Multi-Source Reputation Aggregation}

\author{Nejc Znidar
  \and Anton Kolonin}
\institute{SingularityNET}
\maketitle              

\begin{abstract}
In this paper, we present an extension of liquid rank reputation systems that enables the aggregation and blending of multiple heterogeneous reputation sources into a unified reputation score. The proposed framework supports the incorporation of external reputational signals alongside internally generated reputation, allowing influence to reflect participation and contribution across multiple contexts and subsystems. By introducing explicit weighting and blending mechanisms, the model provides fine-grained control over the relative impact of individual reputation sources, making it adaptable to diverse governance and coordination scenarios involving both human and machine agents. The resulting approach extends existing liquid rank systems and offers a flexible foundation for designing reputation-based governance mechanisms in complex socio-technical environments.
\end{abstract}

\section{Introduction}\label{sec:demonstration}
For most of human history, collective decision-making has been shaped by evolving forms of governance that sought—sometimes successfully, often imperfectly—to align power with legitimacy. Early civilizations relied on hereditary authority or coercive rule, where the right to govern flowed from birth, conquest, or possession of resources. However, long before the emergence of formal states, many small-scale societies also practiced participatory or consensus-based decision-making—forms of early democracy expressed through council deliberation, communal voting, or collective agreement \cite{Service1975Origins}. As cultures grew more complex, these practices developed into more recognizable political systems. Ancient Greece famously formalized democracy as a civic institution, while other regions experimented with monarchy, aristocracy, oligarchy, tyranny, and mixed constitutions, each with differing degrees of stability and inclusiveness.

Even today, human societies exhibit a wide spectrum of governance models, each shaped by historical context, cultural norms, and institutional constraints. Nation-states employ democratic, authoritarian, monarchical, and hybrid systems—some lasting for centuries, others proving far more fragile. Beyond formal governments, smaller communities also rely on governance structures: municipalities operate through councils and public deliberation, while private companies often adopt strongly hierarchical or autocratic decision-making frameworks. Across all these examples, one common pattern emerges: there is no universally perfect model of governance. Every system embodies trade-offs, and even the most stable arrangements eventually evolve, weaken, or collapse.

With the rise of blockchain networks, new forms of digital governance emerged\allowbreak---most notably token-based voting, in which influence is proportional to the number of tokens held. This one-token, one-vote model evolved organically from early crypto-economic systems inspired by shareholder governance in traditional finance, rather than from democratic theory. Its earliest large-scale implementations appeared in governance experiments around Ethereum-based protocols and were later formalized within decentralized autonomous organizations (DAOs), where token ownership became both an economic stake and a proxy of political power. However, this approach exhibits well-known shortcomings: it systematically concentrates influence in the hands of large holders, favors short-term speculation over long-term participation, and often grants governance authority to investors who may have little involvement in the operational or social life of the network \cite{dupont2017dao}. As a result, meaningful contributors—developers, users, maintainers, and community members—remain structurally underrepresented.
This paper focuses on extending the mathematical framework underlying the reputation system originally developed by Kolonin et al. in 2018 \cite{kolonin2018liquidrank}. The existing model has already been validated across a broad range of applications, including blockchain governance systems (e.g., Proof-of-Reputation consensus) \cite{aluko2021proof}, social media analysis \cite{kolonin2024socialmedia}, content recommendation \cite{kolonin2022recommendation}, online community management \cite{kolonin2019onlinecommunities}, and several additional use cases. Building on this foundation, we propose an enhanced framework in which multiple sources of user reputation are merged into a unified, context-specific score, offering a more complete and accurate representation of a participant’s role within a given ecosystem. By incorporating heterogeneous reputation signals into a coherent structure, this approach enables more expressive and adaptable governance mechanisms, where the relative impact of each reputation dimension can be formally defined, weighted, and assessed. In doing so, it expands the applicability of the original model to more complex governance scenarios, including those where influence should derive from meaningful contribution to the community rather than financial stake.

\subsection{Reputation system concept}

Earlier versions of the Liquid Rank reputation system were introduced by Kolonin et al. \cite{kolonin2018liquidrank}, and the present work builds upon and extends this line of research. The principle is called Proof-of-Reputation (PoR). This approach contrasts with alternative governance mechanisms, - most notably those in which voting power is determined primarily by financial stake. In token-based governance models, such as the widely adopted one-token-one-vote scheme in cryptocurrency systems, influence over decision-making is directly proportional to token ownership, leading to a concentration of power among large holders. In Liquid Democracy, the power of a node, that is an entity, be it a human or artificial participant, depends on the reputation of that participant. Reputation is accumulated through "ratings", that is explicit or implicit values attached to real transactions in respect to the accounts given those transactions, called "raters". "Ratees" are the accounts receiving the transaction and reputation. The computation is dependent on time, is updated every period (the definition of period can be chosen by system design, with daily period commonly used as a default) and recent updates have a higher weight than those far in the past. By design, the data used for reputation computation is publicly available, enabling transparency and retrospective analysis of reputation dynamics.

A wide variety of signals may serve as ratings, depending on the objectives of the system. These may include explicit user-provided ratings, upvotes, downvotes, implicit ratings or some other calculations. The specific choice of rating mechanisms is application-dependent. Prior work has demonstrated the applicability of Liquid Rank–based reputation systems in domains such as online marketplaces \cite{kolonin2019reputationmarketplacesviability}, social network analysis \cite{kolonin2024socialmedia}, and related socio-technical environments.

\section{Methodology}

According to Kolonin (2025) \cite{kolonin2025generalizedreputationcomputationontology}, the endorsed liquid rank reputation system operates as follows.
Reputation of rater $i$ in time period $t$ is defined as $R_i(t)$. There is also a default reputation towards which the user's reputation converges assuming there is no activity, called $R^{default}$.
We further introduce endorsing ratings, $S_{ij}$, which represent persistent or structural evaluations issued by agent $j$ to agent $i$, such as endorsements corresponding to relationships like follows or provides. In addition, we define transactional ratings $F_{ije}$, that can be recorded and are associated with either financial transactions (such as pays), or acts of voting in respect to particular events $e(t)$. Those can be publications, posts, comments, tasks, etc. Ratings can be either explicit or implicit. Static ratings $S_{ij}$ can also be backed up by financial stake value or other weight of endorsed rating, $Q_{ij}$. This can be implicit or explicit. The same goes for transactional voting ratings that can be backed by financial value $G_{ije}$. Rating values may be scaled between -1 and 1.
We update those endoresed ratings in the following way:
$$dS_i (t_{n-1}, t_n) = \sum_{k=1}^K * \frac{\sum_{i,j}S_{ij}(t_n)*Q_{ij}(t_n)*R_j(t_{n-1})}{\sum_{i,j}Q_{ij}(t_n)*R_j(t_{n-1})}$$
\label{eq:dst}

where $\sum{H_k}=1, H_k >0, \forall H_k$, $k=1,2,...,K$ where $K$ is the number of different  transactional ratings $k$. 
Similarly, we can get to:
$$dR_i (t_{n-1}, t_n) = \sum_{k=1}^K * \frac{\sum_{i,j}F_{ije}(t_n)*G_{ije}(t_n)*R_j(t_{n-1})}{\sum_{i,j}G_{ije}(t_n)*R_j(t_{n-1})}$$
\label{eq:drt}

This was defined in previous paper \cite{kolonin2025generalizedreputationcomputationontology}. This time blending is adjusted.
Given the normalization of $dR_i$ and $-1<F_{ij}(t_n) < 1$ (and it can also be $dR_i$ and $0<F_{ij}(t_n) < 1$).
Furthermore, the differential reputation update $dR_i$ may be decomposed across multiple reputation types. One can normalize the differential over the types of accounts. Normalization means that among the same reputation types, normalization is done in the following way:
$$dR_i^c = \frac{dR_i^c}{max(dR^c)}$$
where $c$ is the reputation type. An example would be the exchanges - one normalizes differential over exchange type accounts only. This means that there can be multiple accounts with maximum differential - albeit they have different levels of activity. This is because they represent different types of accounts.

All this is done in conjunction with the generalized reputation system, as described in Kolonin (2018) \cite{kolonin2018liquidrank}. Note, that this is what works for endorsed ratings. We can include multiple reputations and endorsed reputation(s) can be counted as one of the external reputations. In the blending section, we then describe how we can merge multiple external reputations.

\subsection{Blending}

In the proposed framework, multiple reputation sources are combined into a unified reputation. Conservatism is retained to ensure gradual rather than abrupt reputation changes, thereby limiting the influence of short-term fluctuations and increasing resistance to reputation manipulation.

Let us define the reputations as $R_i^1,R_i^2,...,R_i^n$. They represent different reputation sources that we merge for the reputation of the agent $i$. One of those is our core reputation that we use. Note that since we use core reputation in default transactions and endorsed ratings, other reputations should be included in the calculations of this reputation. For example, if one of the other reputation sources has a positive reputation in that source, then that adds up to the reputation of the core. We will therefore define the core reputation as $R_{i}^{core}$.

In case we merge multiple reputations into $R_{i}^{core}$, the latter is then calculated slightly differently than others. For other reputations, the computation can be computed similarly in previous papers \cite{kolonin2018liquidrank} - that is without blending or in any specific way that the modeler desires.

$$R_{i}^{core}(t) = (1-C) * (dR_{i}^{core}(t)) + C * R_{i}^{core}(t-1) $$
or
$$R_{i}^{core}(t) = (1 - C) * R_{i}^{core-default} +  C * R_{i}^{core}(t-1) $$
in case that there is no activity. In such a case, the reputation slowly decays toward a default. We also only have a gradual increase in reputation over time.

We next incorporate external reputation sources into the computation of the core reputation. Rather than replacing the internally computed core reputation, each external reputation contributes proportionally according to a predefined blending weight. Let $h_k$, denote the weight associated with reputation source $k$, such that
$$\sum_{k=1}^{n} h_k = 1$$
The blended reputation is defined as
$$ R_i(t) = h_1R_i^{core}(t) + \sum_{k=2}^{n}h_kR_i^{k}(t). $$

The resulting blended reputation $R_i(t)$ is subsequently used as the participant's reputation in all future liquid-rank computations. Consequently, whenever the reputation term $R_j(t-1)$ appears in the differential reputation computation, it refers to the blended reputation defined above.

The differential update of the core reputation is then computed as
$$dR_i^{core} (t_{n-1}, t_n) = \frac{\sum_{i,j}F_{ije}(t_n)*G_{ije}(t_n)*R_j(t_{n-1})}{\sum_{i,j}G_{ije}(t_n)*R_j(t_{n-1})}$$

If an account does not possess a particular external reputation, a predefined default value is assigned for that reputation source. Consequently, all accounts may participate in the blending process regardless of whether they are represented in every external reputation system. Although the framework naturally accommodates external reputations computed using the proposed methodology, this is not a requirement. Any external reputation metric normalized to the interval $[0,1]$, together with an appropriate default value, may be incorporated into the blending mechanism.

\section{Illustrative Behaviour of the Blended Reputation Model}
To illustrate the behavior of the proposed blending mechanism, we consider a simple synthetic example consisting of two reputation sources: the core reputation $R_{core}$ and one external reputation $R_2$. The initial value of both reputations is set to $0.5$, which also corresponds to the default reputation. The blending weights are fixed to:
$C=0.8$,
$h_1 = 0.8$,
$h_2 = 0.2$,
$R_{core}^{default} = 0.5$ and
$R_2 = 0.5$

\begin{figure}
\includegraphics[width=\textwidth]{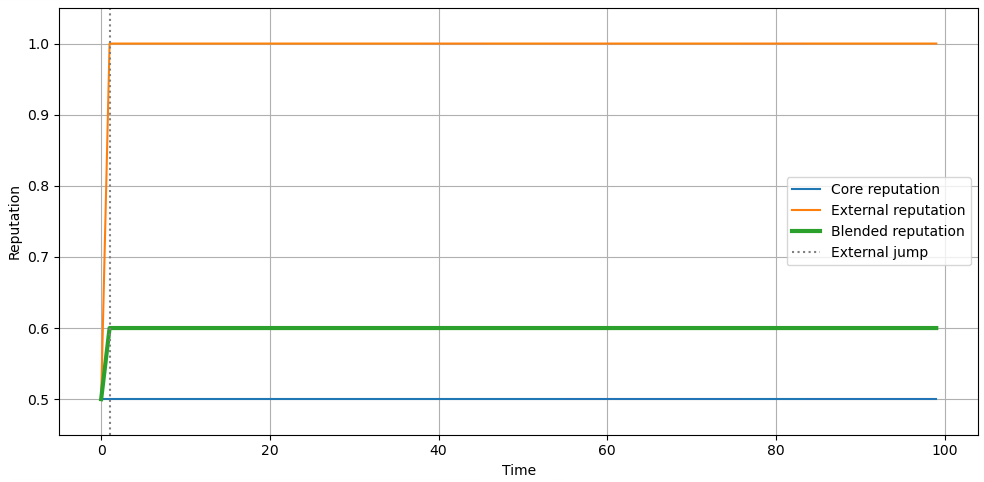}
\caption{Illustration of the blended reputation model without core transactional activity. The external reputation increases from 0.5 to 1.0 after the first update period, causing the blended reputation to increase immediately from 0.5 to 0.6, while the underlying core reputation remains at its default value.} \label{fig:no_activity}
\end{figure}

In the first experiment, the external reputation is assumed to change abruptly from $0.5$ to $1.0$ after the first update period, for example due to successful completion of a KYC verification or another external trust signal. No transactional activity is recorded in the core reputation. Consequently, the core reputation remains equal to its default value, while the blended reputation immediately increases from $0.5$ to
$$0.8 * 0.5 + 0.2 * 1=0.6$$
Figure \ref{fig:no_activity} illustrates this behavior. The external reputation directly affects the blended reputation, while leaving the underlying core reputation unchanged.

\begin{figure}
\includegraphics[width=\textwidth]{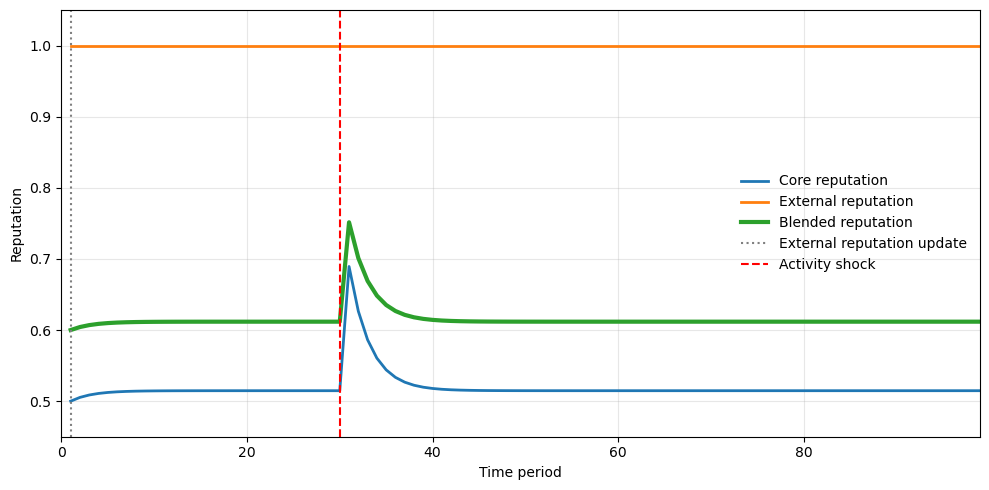}
\caption{Illustration of the blended reputation model under continuous low-intensity transactional activity. The account receives small positive ratings each period, and an additional positive activity shock is introduced at $t=30$. The core reputation responds gradually due to the conservatism parameter, while the blended reputation reflects both the external reputation and the evolving core reputation.} \label{fig:activity}
\end{figure}

In the second experiment, the same external reputation update is applied, but the account additionally receives a small positive transactional reputation signal every period. The differential reputation is computed according to Equation \eqref{eq:drt} using several raters of different reputation levels. Furthermore, at period $t=30$, an additional positive activity shock is introduced by increasing one of the received ratings. The resulting evolution is shown in Figure \ref{fig:activity}.

Initially, the repeated positive ratings gradually increase the core reputation above its default value. The increase remains intentionally small, as the account receives only modest positive ratings from predominantly low- or medium-reputation participants. Consequently, the weighted differential reputation remains close to the default value, illustrating that limited activity or attempts to accumulate reputation through weak reputation sources produce only marginal improvements in the resulting core reputation. Since the blended reputation combines both the external and the core reputation, it follows the same trend while remaining bounded by the blending weights. The temporary activity shock at $t=30$ produces a short-term increase in both the core and blended reputations. Owing to the conservatism parameter, the influence of this event decays gradually over time, and both reputations converge back to their long-run equilibrium in approximately 15 time periods. This illustrates that temporary bursts of activity have only a limited long-term effect, while persistent behavior and external reputation signals determine the steady-state reputation.

\section{Conclusion}
This work has presented a generalized reputation framework for complex governance and coordination systems that enables controlled integration of multiple external reputation sources into a core liquid-rank reputation. The resulting model allows external reputation signals to contribute to the core reputation, enabling participants to accumulate influence not only through participation in the primary system but also through meaningful contributions across associated subsystems. In this way, the framework supports a richer representation of both human and machine behavior and facilitates the construction of socio-technical systems governed by multiple, interacting incentive structures.

A key property of the proposed approach is the explicit controllability of how individual external reputation sources contribute to the overall reputation score. Through appropriate parameter selection, the impact of each source can be precisely weighted, enabling fine-grained alignment between reputation incentives and system objectives. Rather than replacing existing liquid rank reputation systems, the framework extends and complements them by providing a principled mechanism for multi-source reputation aggregation. The proposed framework preserves the desirable properties of liquid-rank reputation systems while extending them to heterogeneous reputation ecosystems in which trust may originate from multiple independent sources.

While the model establishes a flexible theoretical foundation, further work is required prior to real-world deployment. In particular, simulation-based evaluation under different parameter configurations and environmental conditions is needed to assess convergence properties, system dynamics, stability, and resistance to strategic manipulation.

\bibliographystyle{splncs04}
\bibliography{references}

\end{document}